\documentclass{svproc}
\usepackage{graphicx}
\usepackage{amssymb}
\PassOptionsToPackage{hyphens}{url}
\usepackage{microtype}

\usepackage{xcolor}
\usepackage{tikz,xcolor,hyperref}

\definecolor{lime}{HTML}{A6CE39}
\DeclareRobustCommand{\orcidicon}{
	\begin{tikzpicture}
	\draw[lime, fill=lime] (0,0) 
	circle [radius=0.16] 
	node[white] {{\fontfamily{qag}\selectfont \tiny ID}};
	\draw[white, fill=white] (-0.0625,0.095) 
	circle [radius=0.007];
	\end{tikzpicture}
	\hspace{-2mm}
}

\foreach \x in {A, ..., Z}{%
	\expandafter\xdef\csname orcid\x\endcsname{\noexpand\href{https://orcid.org/\csname orcidauthor\x\endcsname}{\noexpand\orcidicon}}
}

\begin{document}
\mainmatter             
\title{An Automated SQL Query Grading System Using An Attention-Based Convolutional\\ Neural Network}
\titlerunning{An Automated SQL Query Grading System Using Self-Attention and CNNs}

\author{Donald R. Schwartz\inst{1}\orcidA{}
\and Pablo Rivas\inst{2}\orcidB{}
}
\authorrunning{D. Schwartz and P. Rivas} 
\tocauthor{
D. Schwartz, P. Rivas
}

\institute{
$^1$ School of Computer Science and Mathematics\\
Department of Computing Technology \\
Marist College, New York, USA\\
\email{Donald.Schwartz@Marist.edu}\\
$^2$ School of Engineering and Computer Science \\
Department of Computer Science \\
Baylor University, Texas, USA\\
\email{Pablo\_Rivas@Baylor.edu}
}

\maketitle

\begin{abstract}
Grading SQL queries can be a time-consuming, tedious and challenging task, especially as the number of student submissions increases. Several systems have been introduced in an attempt to mitigate these challenges, but those systems have their own limitations. This paper describes our novel approach to automating the process of grading SQL queries.  Unlike previous approaches, we employ a unique convolutional neural network architecture that employs a parameter-sharing approach for different machine learning tasks that enables the architecture to induce different knowledge representations of the data to increase its potential for understanding SQL statements.

\end{abstract}

\section{Introduction \label{sec:int}}

One of the common challenges when grading SQL queries is the large number of possible correct answers.  For example, a simple query such as “Name the professors who teach a class called ‘Introduction to Programming’” could be written very many different ways:  as a non-nested query with three tables in the {\tt FROM} clause, as a nested {\tt EXISTS} query, as a nested IN query, as a nested {\tt IN/EXISTS} query, as a {\tt NATURAL JOIN}, as a {\tt JOIN ON}, as a {\tt JOIN USING}, with or without aliasing table names, with or without aliasing field names in the answer table, etc., etc., etc.  This challenge significantly increases as the complexity of the queries increases. Another challenge arises when considering partial credit for queries that are not correct.  Being consistent with partial credit given the wide range of possible answers is very important but is not always easy. And a significant challenge is simply the shear tedium of grading dozens of queries for scores of students over the course of the semester!

This paper describes our attempt to help mitigate these challenges by automating the grading process.  We begin with background material, describing various approaches to automating the grading process, convolutional neural networks, and self-attention mechanisms.  Next, we describe the database components within our system, the machine learning system we have developed and the results of our experiments. We conclude with the benefits of our system and future work.

\section{Background}

Most automated grading systems fall into one of two categories:  static analysis systems and dynamic analysis systems.

Static analysis systems generally evaluate SQL queries without actually running them.  They do so by comparing the structure of a submitted query with the structure of an answer query. An early example of this approach~\cite{aho1979equivalences} used a matrix to represent a variety of relational expressions in order to establish equivalence classes for those expressions. Another system developed by Štajduhar et al. compares queries by employing string similarity metrics~\cite{vstajduhar2015using}. Cosette is a system developed by Chu et al. that determines whether queries are logically equivalent by encoding the queries into logical expressions~\cite{chu2017demonstration}.  They advanced this approach by later developing an approach using an unbound semiring to determine the level of equivalence of queries~\cite{chu2018axiomatic}. This improved system allowed for more detailed feedback to the students but was a much more complex approach. It required all possible correct answers to be included in the answer key, so that the system could compare the logical expression for the student’s query with the logical expressions of all possible answers in order to award the student with the highest possible resulting score.

The other approach is dynamic analysis, in which submitted queries are actually run against one or more datasets and the resulting table is compared to an answer key~\cite{wang2020combining}. There are many early examples of this approach, including AsseSQL~\cite{prior2004backwash}, SQLify~\cite{dekeyser2007computer} and SQLator~\cite{sadiq2004sqlator}.  SQL Tester is an online tool developed by Kleerekoper et al.  This interactive tool does a case-sensitive comparison of the resulting tables, which requires the records to appear in the same order~\cite{kleerekoper2018sql}. ASQLAG is a system capable of evaluating queries across a variety of database platforms (including SQL Server, PostgreSQL, and MySQL) by using an object-oriented design approach employing a Model-View-Controller (MVC) framework~\cite{singporn2018asqlag}. Systems that compare the answer table to the submitted table often cannot accurately determine whether the query itself is correct.  For example, in the query described above, it might be the case that in the current state of the database, the Professors whose ID number is between 1000 and 2000 might be the ones teaching Introduction to Programming.  A query based on ID numbers should certainly not earn credit for the original query. This weakness is especially apparent for queries whose answer is an empty table.  However, comparing the answer table to the submitted table can certainly identify those queries thar are NOT correct.  This information can be very useful, but such systems are not well-equipped to accurately assign partial credit to those queries. An example of this is when a submitted query results in one extra (or one fewer) field than the answer key.

Some more-recent systems combine the two approaches.  Wang et al. have developed a system that identifies incorrect queries by using dynamic analysis, then employs static analysis on those queries to determine how much partial credit to award. Partial credit is determined by comparing the submitted query to multiple correct answers, awarding the highest resulting value~\cite{wang2020combining}.  The XData system, developed by Chandra et al., takes a similar approach.  This system dynamically analyzes queries by generating specific datasets designed to expose common errors students make when developing those queries.  It then uses static analysis to compute how many changes must be made to the submitted query in order to transform it to an equivalent correct query~\cite{chandra2021edit,chandra2019automated,chandra2016partial}.

With the introduction of attention mechanisms in 2017 by Vaswani et al., researchers 
have developed robust algorithms to attend to specific portions of sequential data for classification, regression, and other machine learning tasks~\cite{vaswani2017attention}. These include novel architectures, such as combining an attention mechanism with a convolutional neural network (CNN), in applications in the domain of medical imaging~\cite{wu2019self,li2020sacnn}, sequential data~\cite{fahim2020self}, and others~\cite{zeng2020crop}. In all cases, the data contains correlated information across a temporal feature space. Researchers that used these architectures have reported  improvements over classic approaches.

The success of these kinds of architectures is that CNNs have an inductive bias toward highly correlated data in a defined n-dimensional space, including images, time series, sentences, or SQL statements. The self-attention mechanism provides contextual information about the data within its context, making the architecture very powerful. 

\begin{figure}[t]\centering
\begin{center}
\includegraphics[width=\textwidth]{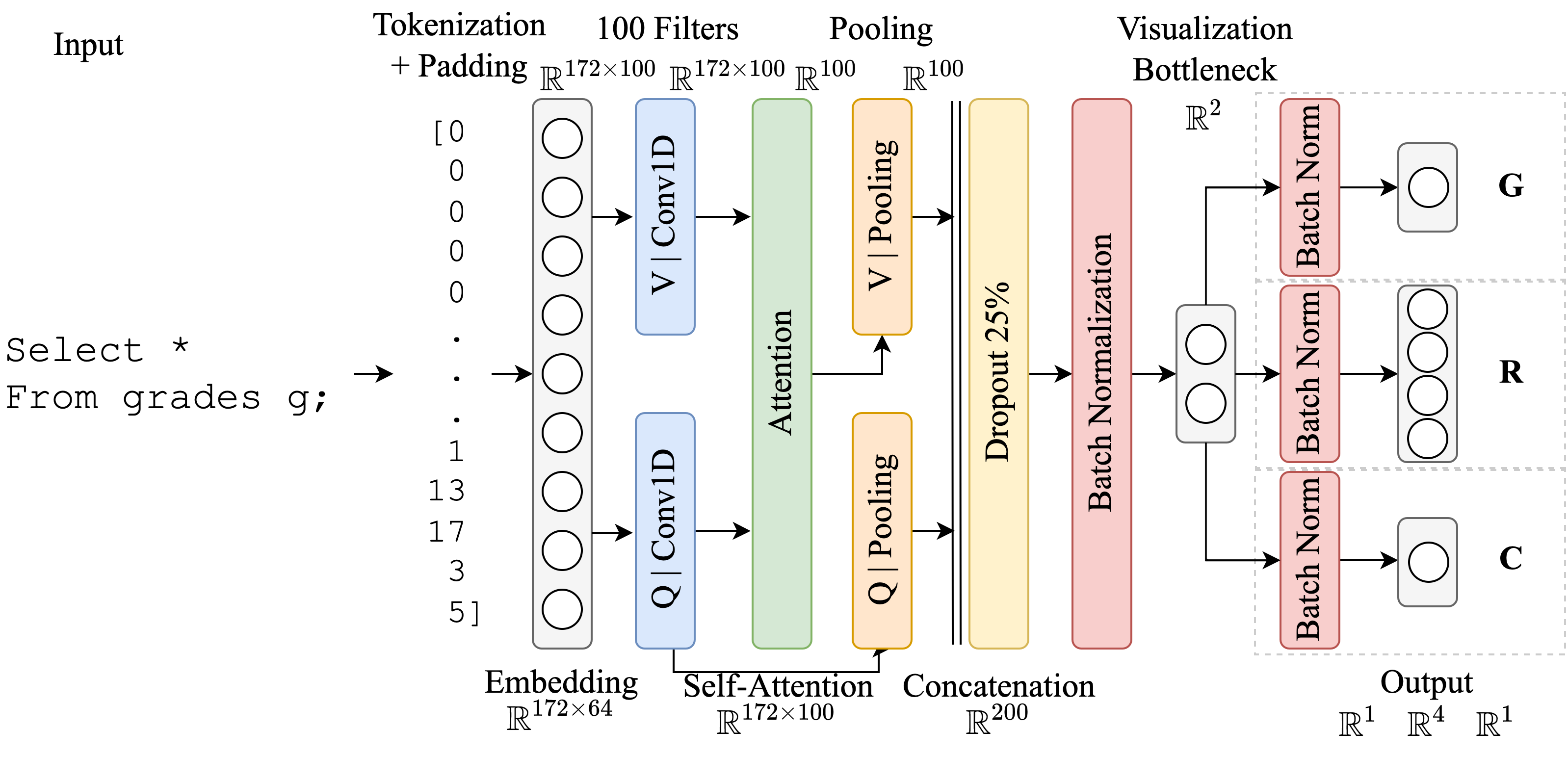}
\vskip -0.2in
\caption{Proposed neural architecture for SQL Statement grading based on self-attention, CNNs, and parameter sharing.
\label{fig:architecture}}
\end{center}\vskip -0.2in
\end{figure}

Other work has been done in modeling SQL statements~\cite{chen2021shadowgnn,bogin2019representing}; however, the authors have significantly different data, aims, and goals. Architecturally, our proposed model, depicted in Fig. \ref{fig:architecture}, is different as it uses a parameter-sharing approach for different machine learning tasks that enable the architecture to induce different knowledge representations of the data to increase its potential for SQL statement understanding. To the best of our knowledge, no other approaches have used this architecture and learning paradigms on SQL statement grading.

\section{Description of the Database Components}

Our overall database contains all of the tables required for the assignments (which, using the above example, would include at least {\tt PROFESSORS, TEACHES} and {\tt COURSES}), along with the following tables, some of which are used to provide the information needed for our proposed system.

The {\tt STUDENT} table contains identification information about each student registered for each section.

The {\tt QUERY} table consists of the {\tt QueryID}, the text/wording of the query, and the number of possible points for that query.

The {\tt ASSIGNMENT} table describes which queries are part of which lab assignment.

The {\tt ANSWER} table holds the {\tt AnswerID}, the {\tt QueryID}, and a correct SQL answer for that query.  Naturally, there are multiple {\tt AnswerID}s associated with each {\tt QueryID}.

The largest table is the {\tt STUDENTSUBMISSION} table.  This table holds the {\tt StudentID, QueryID, SubmittedAnswer} (the SQL query submitted by that student), a flag indicating whether the {\tt SubmittedAnswer IsCorrect}, a {\tt Remark} field that describes whether the submission was {\tt correct, partially correct, uninterpretable}, and the number of {\tt Points} earned.

We populated our database with queries written by our students over the past couple of years. This required us to regrade the assignments to remove penalties that didn’t necessarily affect the queries themselves, for example, late penalties, deductions for not following directions, and penalties for using non-descriptive aliases for tables and fields.  Our datasets included roughly 2700 individual student queries.

The following data is fed into our system: {\tt SubmissionID} (which is based on the {\tt studentID} and {\tt queryID}), {\tt SubmittedAnswer, IsCorrect, Remark} and {\tt Grade} (the number of points earned divided by the possible points for that query).

\section{Description of the Machine Learning System}
The neural architecture displayed in Fig. \ref{fig:architecture} will be discussed in detail next.

\subsection{Embedding Layer}
The embedding layer is a trainable set of neurons that take a token in the context of the entire sequence~\cite{ghazanfari2020multi,rivas2020deep}. Our sequence has size 172 and returns a vector of size 64 to represent a token in the sequence.

\subsection{Convolutional Self-Attention Layer}
The self-attention CNN is composed of three major parts. First, we have two convolutional encoding layers that model a query Q and value V. These are classic elements of attention mechanisms~\cite{vaswani2017attention}. For self-attention, the input to both Q and V is the same embedding. Second, the self-attention layer uses a dot product similarity~\cite{luong2015effective}.  Third, there is a pooling strategy that implements a global average, reducing the dimensionality of the problem down to two vectors of size 100 each~\cite{li2019teeth}. During training, this layer learns optimal embeddings based on self-attention.

\subsection{Dropout Layer}
To regularize the model and reduce the chances of overfitting we used a 25\% dropout~\cite{srivastava2014dropout}. This mechanism further promotes a holistic learning from the attention embeddings themselves. 

\subsection{Batch normalization Layers}
To improve numerical stability during learning and help regularize the learning problem, we used batch normalization layers. Batch normalization calculates optimal means and variances for data batches reducing, vanishing or exploding gradient problems~\cite{santurkar2018does}.

\subsection{Bottleneck Layer}
The model has a bottleneck dense layer with hyperbolic tangent activations that enables us to visualize the learned representation space. We use two neurons only, allowing us to display the learned representations in two dimensions if needed.

\subsection{Outputs C, R, and G}
The right-most portion of Fig. \ref{fig:architecture} displays three different groups of outputs, each of which are different models that share the same parameters of the prior layers, up to the bottleneck. These models each have a batch normalization layer and dense neural units. Model \textbf{C} is for predicting correctness, model \textbf{R} is for predicting the grader's remark, and model \textbf{G} is for predicting the grade. Each of these models is trained iteratively as we discuss in the next section.

\subsection{Experiments, Evaluation, and Results}
The model in Fig. \ref{fig:architecture} was implemented in Python using the Tensorflow platform with Keras libraries. The training discussed below was performed on an NVIDIA P100 GPU system with 25 GB of RAM and 166 GB of storage.

We performed two major experiments: 1) a jointly trained model with the purpose of validating correctness on k-fold cross-validation; and 2) a jointly trained model that validates all models using leave-one-out cross-validation. 

For our experiments the three models \textbf{C}, \textbf{R}, and \textbf{G} are trained one at a time with different loss functions optimized using gradient descent with an RMSprop optimizer and learning rate of 0.001~\cite{mukkamala2017variants}.

\subsubsection{Jointly trained model cross-validated performance.}
During training in this experiment, the gradient is calculated using a single loss: the binary cross entropy loss. The length of the output is a vector of six elements: one item for the correctness, four items for the remarks, and one item for the grade. We measure the performance using the area under the receiving operating characteristic (ROC) curve (AUC) using 10-fold cross validation. The motivation for measuring this at this point is to make an early assessment before launching a full-scale performance evaluation.

\begin{figure}[t]\centering
\begin{center}
\includegraphics[width=0.95\textwidth]{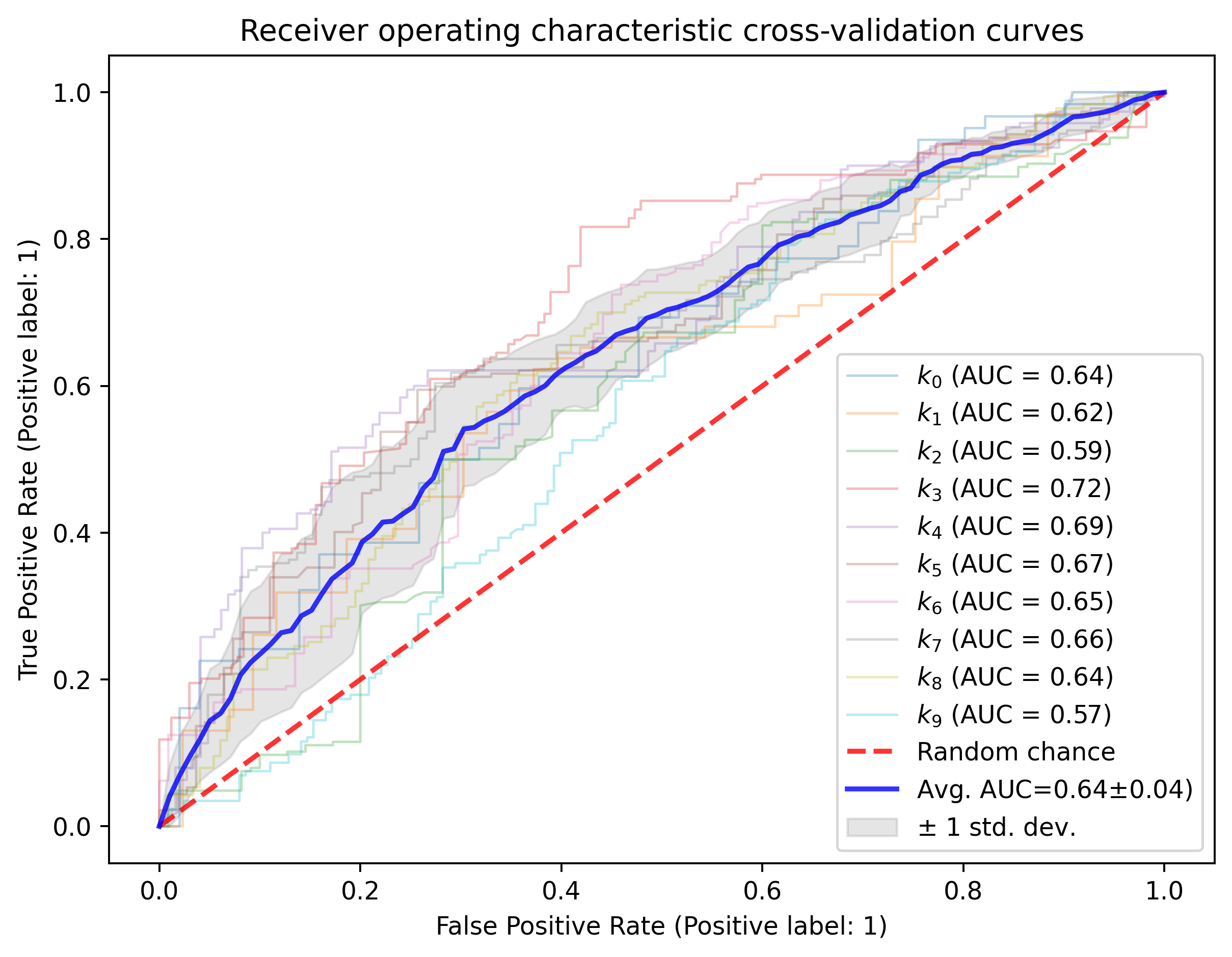}
\vskip -0.2in
\caption{ROC curves and corresponding AUC across the 10 different folds. The average AUC is 0.64. In all folds, the model performs better than random chance.
\label{fig:aucs}}
\end{center}\vskip -0.2in
\end{figure}

The results of using cross validation are displayed in Fig. \ref{fig:aucs}. From the figure we can assert that the model, in average across folds, performs better than random chance in all cases. The maximum AUC reported was 0.72 and the minimum was 0.57, with an average of \textbf{0.64} and very low variance, 0.04. To improve the estimation of the generalization error we also performed a full cross-validation analysis using a leave-one-out (LOO) strategy across all models.

\begin{figure}[t!]\centering
\begin{center}
\includegraphics[width=0.6\textwidth]{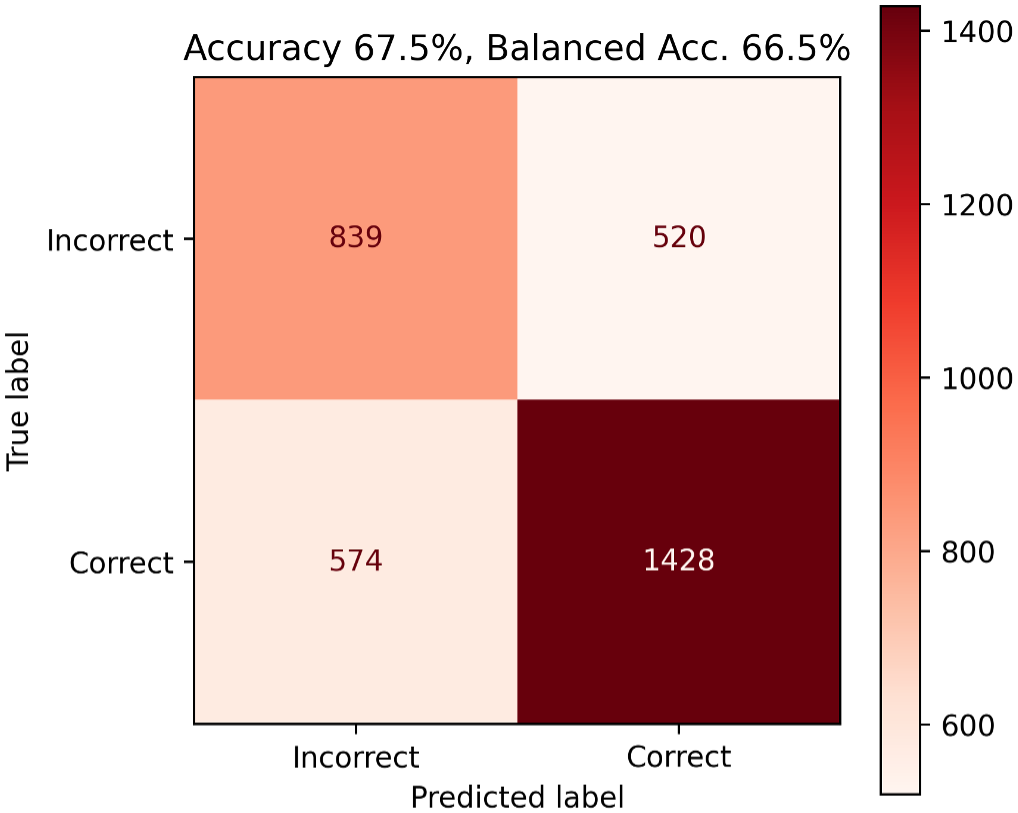}
\vskip -0.2in
\caption{Confusion matrix for the model that predicts correctness.
\label{fig:cm}}
\end{center}\vskip -0.2in
\end{figure}

\begin{figure}[t!]\centering
\begin{center}
\includegraphics[width=0.9\textwidth]{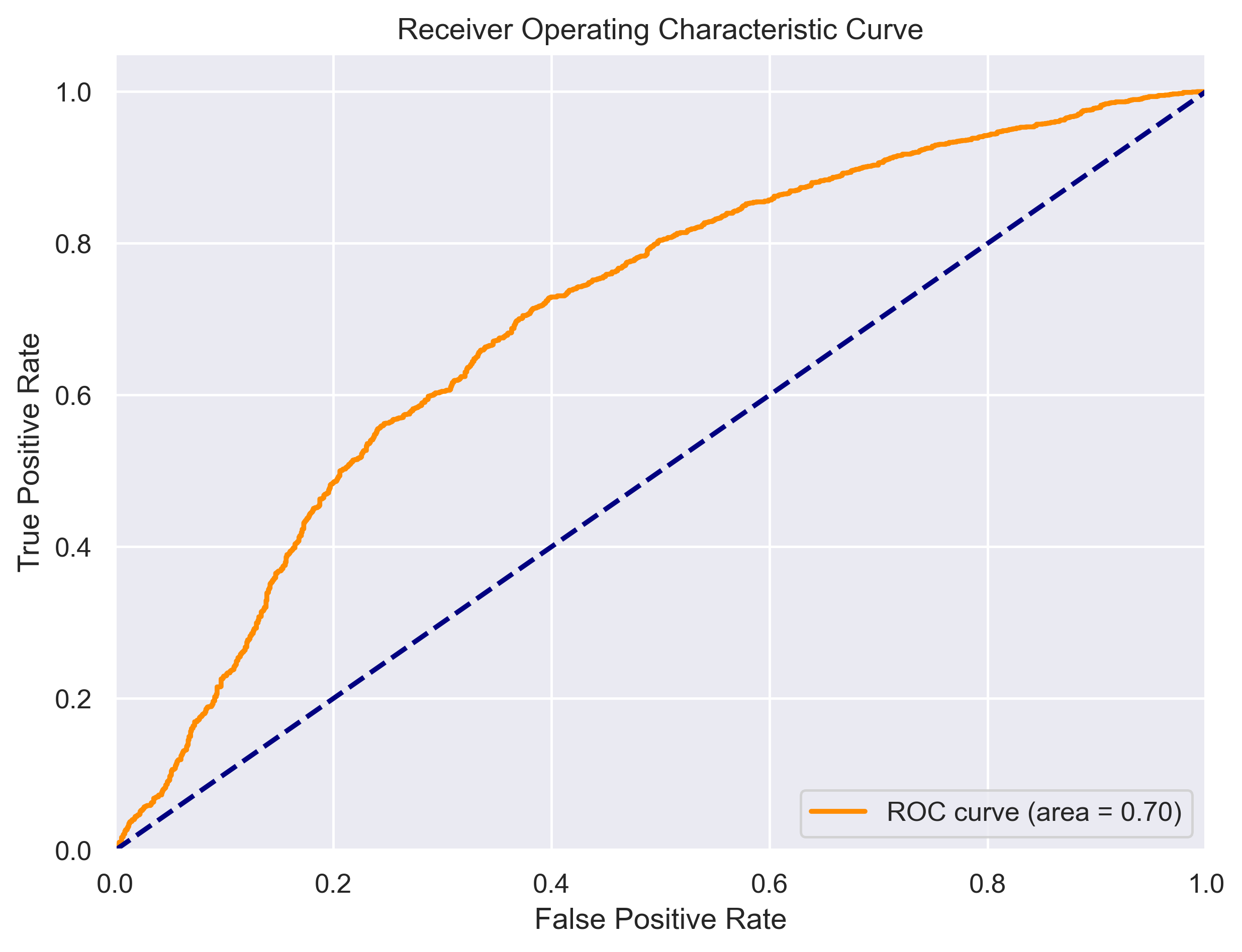}
\vskip -0.2in
\caption{ROC and AUC for the model that predicts correctness.
\label{fig:roc}}
\end{center}\vskip -0.2in
\end{figure}

\subsubsection{LOO cross-validation on all models.}
LOO cross-validation is well-known as the most robust generalization error estimator, particularly when it comes to small datasets such as the one we use~\cite{kearns1999algorithmic,wang_jinshui_2020_6526769}. For the correctness model, \textbf{C}, after training under a LOO analysis, we obtain the results shown in Fig. \ref{fig:cm} and \ref{fig:roc}. The confusion matrix shown in Fig. \ref{fig:cm} indicates the slightly higher false negative (FN) count on the correctness predictions; however, the class imbalance of the dataset may be affecting this outcome. This imbalance obligates us to consider the balanced accuracy instead of the traditional accuracy score. In this case the balanced accuracy is 66.5\%.
We also calculate the LOO cross-validated ROC and AUC, shown in Fig. \ref{fig:roc}. From the figure we observe that the model performs far beyond random chance and reaches an AUC of \textbf{0.70}.

Now we briefly look at mistakes, specifically beginning with an example of the \emph{best mistakes}, i.e., mistakes that were so close to the actual prediction but did not make the classification threshold by two or three decimal points, or in other words, mistakes with high uncertainty. The \emph{best mistake} was:
\begin{verbatim}
SELECT id, first_name,last_name
FROM person p
WHERE p.year_born = (SELECT MIN(p1.year_born)
      FROM person p1);
\end{verbatim}

The sample above is labeled as correct=0, also remark= partially correct, and with grade=40. However, our model predicted it incorrectly as a correct=1 SQL statement with the highest uncertainty, which can be explained by the grade and partial credit given. 

Now let’s inspect the \emph{worst mistake}, i.e., a mistake that was made with highest confidence. Such a sample is:
\begin{verbatim}
SELECT DISTINCT ztv.show_name, SUM(zprb.prod_salary) + 
SUM(zply.actor_salary) AS 
Total_Price
FROM ztvshow ztv, zprodby zprb, zplay zply, zactor zact
WHERE zprb.show_num = ztv.show_num
AND zply.show_num = ztv.show_num
AND zact.actor_num = zply.actor_num
GROUP BY ztv.show_name;
\end{verbatim}

The sample above is labeled as correct=0, also remark=partially correct, and with grade=66.66. Once again this was predicted incorrectly as a correct=1 but this time with the highest degree of certainty. Again, notice the higher grade that it receives and that it is partially correct, regardless of the lack of aesthetics in the SQL statement. 

\begin{figure}[t!]\centering
\begin{center}
\includegraphics[width=0.9\textwidth]{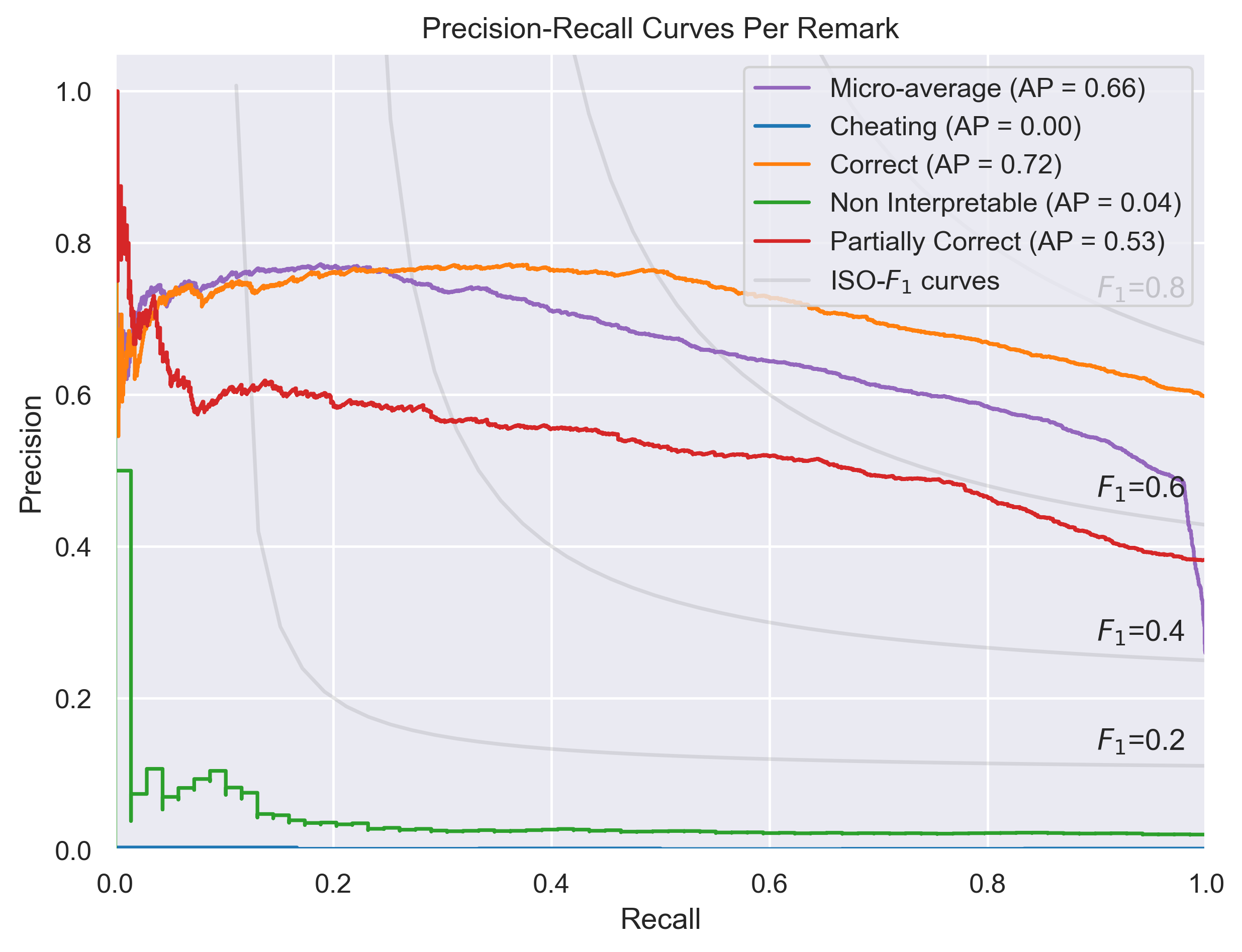}
\vskip -0.2in
\caption{Precision-Recall curves and average precision scores per remark.
\label{fig:prcs}}
\end{center}\vskip -0.2in
\end{figure}

\begin{figure}[t!]\centering
\begin{center}
\includegraphics[width=0.6\textwidth]{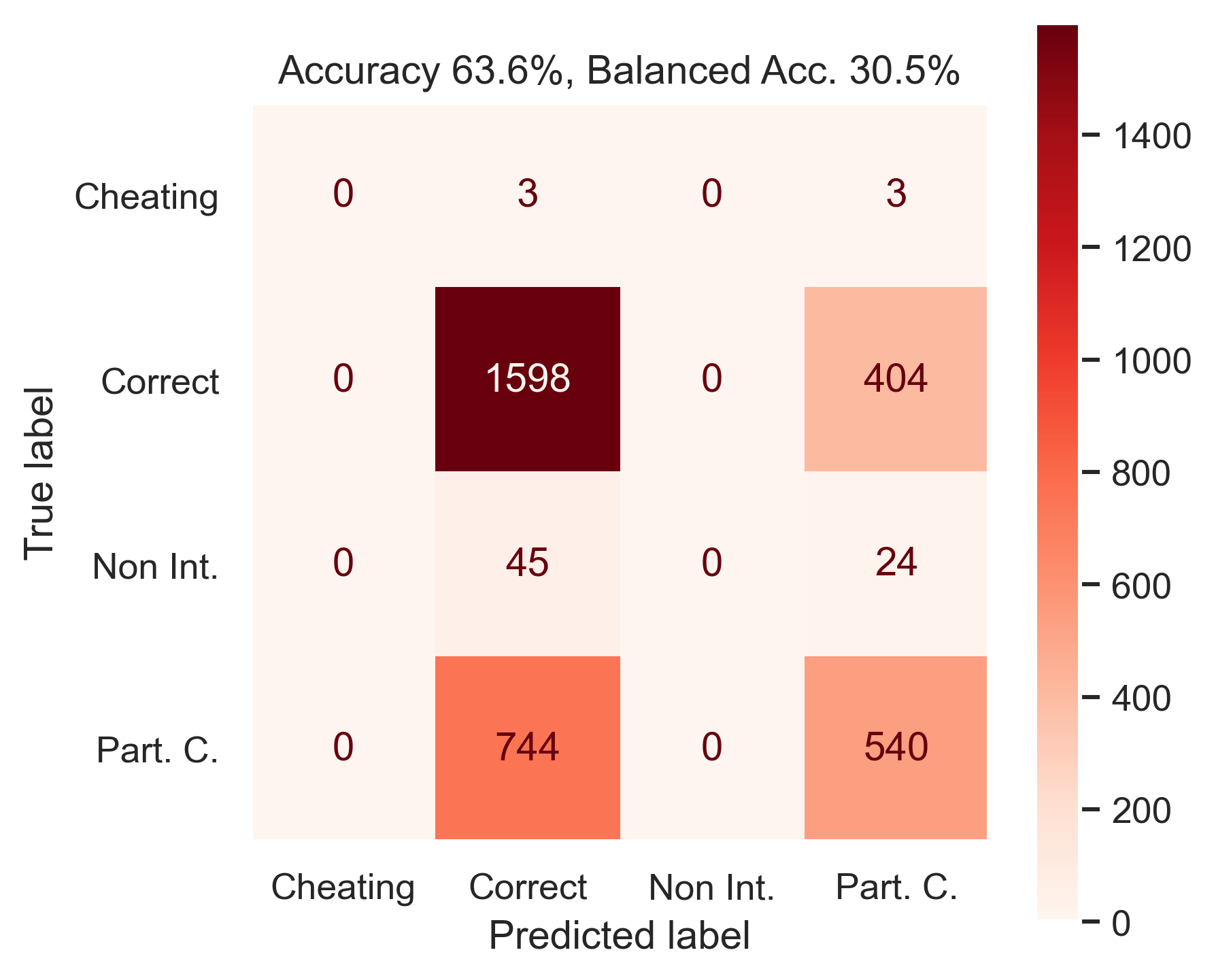}
\vskip -0.2in
\caption{Confusion matrix for the model that predicts remarks.
\label{fig:rcm}}
\end{center}\vskip -0.2in
\end{figure}

For the remarks model, \textbf{R}, after performing LOO, we obtain the results shown in Fig. \ref{fig:prcs}. The figure shows that the model associated with the remarks about correctness, i.e., Correct and Partially Correct, exhibit better performance in comparison to the remarks associated with negative feedback. These results are consistent with our previous finding that correctness can be better predicted than non-correctness. In Fig. \ref{fig:prcs}, for reference, the average precision (AP) score is listed, which is a weighted mean of precisions; it also shows the $F_1$ curves for reference. In Fig. \ref{fig:prcs} and the confusion matrix in Fig. \ref{fig:rcm} we can also observe that the lowest-performing class was Cheating; this is due to the limited number of samples available.

\begin{figure}[t!]\centering
\begin{center}
\includegraphics[width=0.85\textwidth]{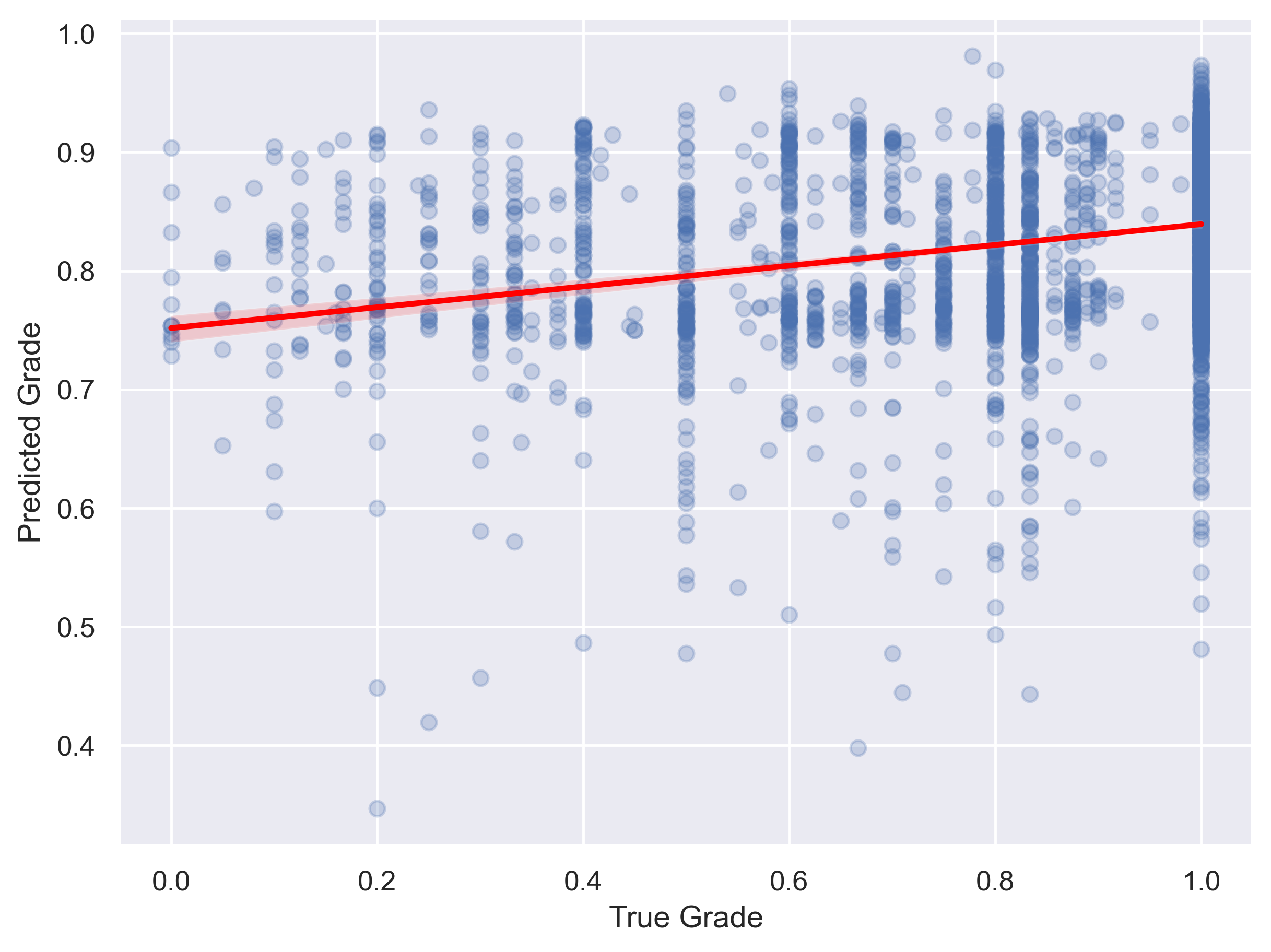}
\vskip -0.2in
\caption{True vs predicted regression plot for the model that predicts grades.
\label{fig:reg}}
\end{center}\vskip -0.2in
\end{figure}

\begin{figure}[t!]\centering
\begin{center}
\includegraphics[width=0.85\textwidth]{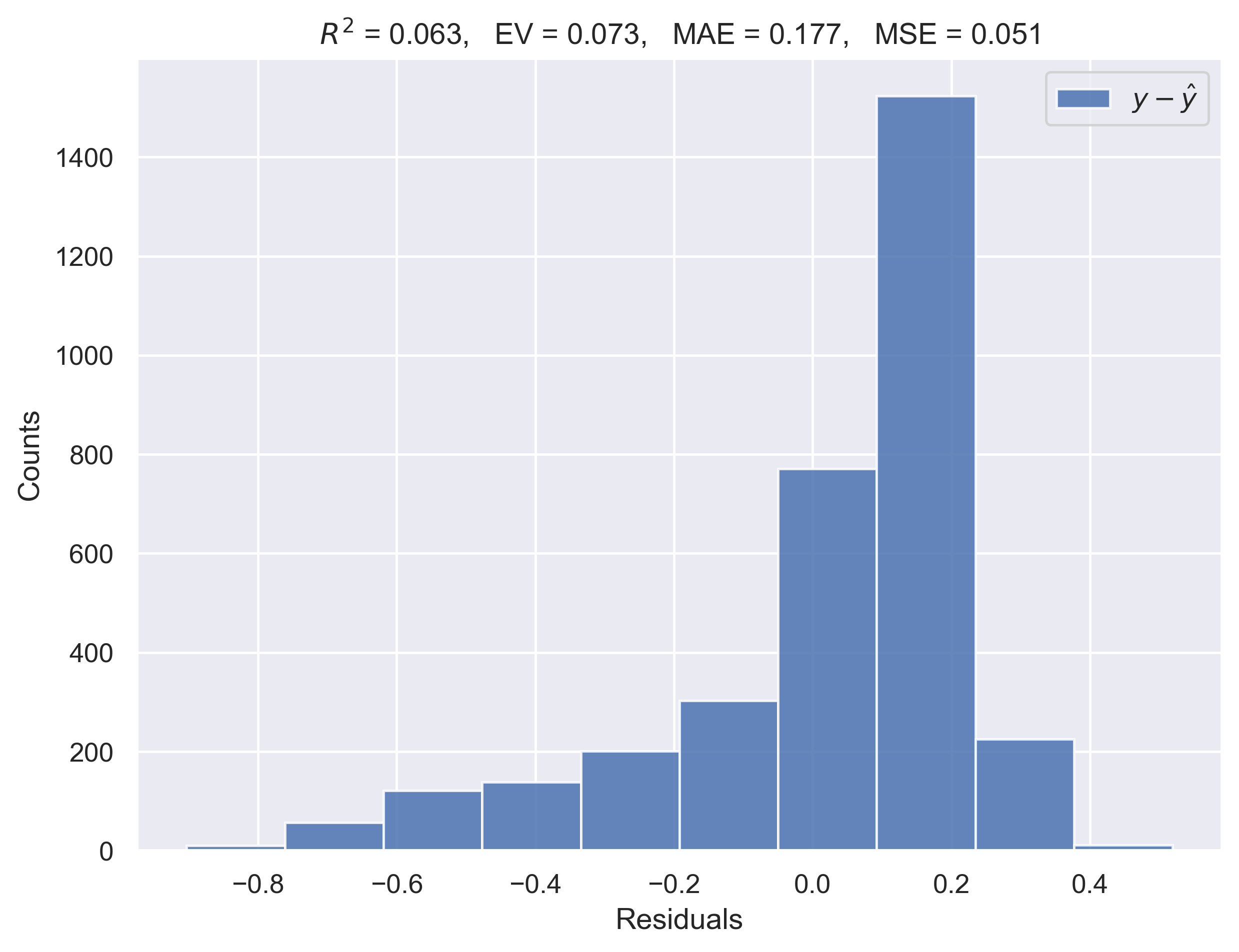}
\vskip -0.2in
\caption{Histogram of the residuals $y-\hat{y}$ for the model that predicts grades.
\label{fig:res}}
\end{center}\vskip -0.2in
\end{figure}

Finally, for the grades model, \textbf{G}, the LOO cross-validated performance analysis can be summarized in Fig. \ref{fig:reg} and \ref{fig:res}. A regression comparison between the predicted grade and the actual grate suggests that our model disfavors grades that are exactly 100\% and favors a more conservative prediction in a wider range. This can be seen in the histogram of the residuals in Fig. \ref{fig:res}, which shows a slight positive skew, while the ideal residuals should be centered around zero. However, the regression performance metrics still capture the predictive ability of the model, which is beyond simply predicting the expected value of the  variable Grade.

\section{Conclusions and Future Work}

As machine learning-based language modeling applications continue to increase in number and variety, we anticipate more research in the general areas of SQL modeling for different analysis and automated decision-making tasks. In this paper we presented a light-weight language model based on convolutional self-attention and parameter-sharing trained across different tasks on SQL statement data. The presented model can be used within another system that processes SQL-based assignment analysis as downstream tasks. We believe this type of system can effectively address the problem of automatic grading of SQL statements. 

Future research will experiment with large language models for an assessment of performance as a function of hyperparameter size; further, the collection of more labeled data with a wider variety of submissions and assignments can be analyzed to distinguish differences in performance according to assignment types, e.g., assignments that vary in difficulty and performance analysis by difficulty.

\section*{Acknowledgements}
The ML model is based upon work supported in part by the National Science Foundation under Grants CHE-1905043, CNS-2136961, and CNS-2210091.

\bibliography{references}

\end{document}